\documentclass[10pt,aps,twocolumn,prl,superscriptaddress,showpacs,nofootinbib,noshowkeys,floatfix,preprintnumbers]{revtex4-1}
\usepackage[dvipdfmx]{graphicx}
\usepackage[usenames]{color}
\usepackage{amsmath,amssymb}
\usepackage{multirow}
\usepackage{longtable}
\usepackage[normalem]{ulem}
\usepackage{epstopdf}
\usepackage{times}
\usepackage[normalem]{ulem}  

\renewcommand\sout{\bgroup \color{red} \ULdepth=-.5ex \ULset}
\usepackage{bm}
\graphicspath{{./Figs/}}
\begin{document}
\title{$K^-p$ Correlation Function from High-Energy Nuclear Collisions
and Chiral SU(3) Dynamics}
\date{\today}
\author{Yuki Kamiya}
\email{yukikamiya@mail.itp.ac.cn}
\affiliation{
CAS Key Laboratory of Theoretical Physics, Institute of Theoretical Physics,
Chinese Academy of Sciences, 
Beijing 100190, China}
\author{Tetsuo Hyodo}
\affiliation{Yukawa Institute for Theoretical Physics, Kyoto University,
Kyoto 606-8502, Japan}
\affiliation{Department of Physics, Tokyo Metropolitan University,
Hachioji 192-0397, Japan}
\author{Kenji Morita}
\affiliation{RIKEN Nishina Center, Wako 351-0198, Japan}
\affiliation{National Institutes for Quantum and Radiological Science and Technology, Rokkasho Fusion Institute, Rokkasho, Aomori, 039-3212, Japan}
\author{Akira Ohnishi}
\affiliation{Yukawa Institute for Theoretical Physics, Kyoto University,
Kyoto 606-8502, Japan}
\author{Wolfram Weise}
\affiliation{Physics Department, Technical University of Munich, D-85748 Garching, Germany}
\affiliation{ExtreMe Matter Institute (EMMI) at GSI, D-64291 Darmstadt, Germany}
\preprint{YITP-19-99}

\begin{abstract}
The two-particle momentum correlation function of a $K^-p$ pair
from high-energy nuclear collisions is evaluated
in the $\bar{K}N$-$\pi\Sigma$-$\pi\Lambda$ coupled-channel framework.
The effects of all coupled channels together with the Coulomb potential and the threshold energy difference between $K^-p$ and $\bar{K}^0 n$ are treated completely for the first time. Realistic potentials based on the chiral SU(3) dynamics are used which fit the available scattering data. 
The recently measured correlation function is found to be well reproduced
by allowing variations of the source size 
and the relative weight of the source function of $\pi\Sigma$ 
with respect to that of $\bar{K}N$.
The predicted $K^-p$ correlation function from larger systems indicates that the investigation of its source size dependence is useful in providing further constraints in the study of the $\bar{K}N$ interaction.
\end{abstract}
\maketitle

\textit{Introduction:}
Low-energy properties of the strong interaction are governed by the symmetry breaking pattern of quantum chromodynamics (QCD). In this context the antikaon ($\bar{K}\sim s\bar{u}/s\bar{d}$) can be regarded as a Nambu-Goldstone boson associated with the spontaneous breaking of three-flavor chiral symmetry. However, its mass is more than three times heavier than that of the pion. Thus, studies involving the antikaon reflect the interplay between spontaneous and explicit breakings of chiral symmetry in low-energy QCD.

The antikaon-nucleon ($\bar{K}N$) interaction at low energy is  strongly attractive. It is the main ingredient to generate the $\Lambda(1405)$ resonance as a $\bar{K}N$ quasibound state~\cite{Dalitz}.
This observation inspired an intense discussion of possible $\bar{K}$-nuclear quasibound systems~\cite{Akaishi:2002bg}. A possible candidate is recently reported by the J-PARC E15 Collaboration~\cite{Ajimura:2018iyx}.

Contrary to its importance in hadron physics
and also in nuclear many-body problems with strangeness,
empirical information on the low-energy $\bar{K}N$ interaction is quite limited.
The $K^{-}p$ scattering amplitude close to threshold 
is accurately constrained by the measurement of the atomic energy shift and width
of $K^-$ hydrogen~%
\cite{Bazzi}.
There exist several $K^-p$ cross section data at relatively high momenta,
$p_\mathrm{Lab}(K^-) > 100~\mathrm{MeV}/c$~\cite{Abrams:1965zz,Sakitt:1965kh,Kim:1965zzd,Csejthey-Barth:1965izu,Mast:1975pv,Bangerter:1980px,Ciborowski:1982et,Evans:1983hz}.
However, the uncertainties of the cross sections from the bubble chamber measurements are large, 
and almost no data exist in the low-momentum region, 
$p_\mathrm{Lab}(K^-) \leq 100~\mathrm{MeV}/c$. 

One of the promising observables that provides stronger constraints 
on the $\bar{K}N$ interaction is the two-particle $K^-p$ momentum correlation
function. It is defined as the two-particle production probability
normalized by the product of single-particle production probabilities~\cite{Koonin:1977fh,Cho:2017dcy}.
Theoretically, the correlation function reflects the two-body interactions
through the wave function with a suitable boundary condition. On the experimental side,
correlation functions have been 
measured recently
in high-energy nuclear collisions
for $p\Lambda$~\cite{Adams:2005ws,Acharya:2018gyz},
$\Lambda\Lambda$~\cite{Adamczyk:2014vca,Acharya:2018gyz},
$p\Xi^-$~\cite{Acharya:2019sms},
$p\Omega$~\cite{STAR:2018uho},
$pK^-$~\cite{Acharya:2019bsa},
and $p\Sigma^0$~\cite{Acharya:2019kqn}
pairs. These data have been used to constrain the pairwise interactions~\cite{Morita:2014kza,Ohnishi:2016elb,Cho:2017dcy,Hatsuda:2017uxk,Haidenbauer:2018jvl,Morita:2019rph}.
Recently the $K^-p$ correlation function has been extracted
from high-multiplicity events of $pp$ collisions~\cite{Acharya:2019bsa}.
The precision of these data is such that even the
$\bar{K}^{0}n$ threshold cusp is visible. 
In addition the data show a peak in the energy region of the  $\Lambda(1520)$ resonance which couples to the $\bar{K}N$  $d$-wave.

For the detailed analysis of the $K^-p$ strong interaction in comparison with the high-precision data it is mandatory to
include the coupled-channel effect, the Coulomb interaction
in the $K^{-}p$ system, and the threshold energy differences among
the isospin multiplets in calculating the correlation function.
While theoretical studies of the $K^-p$ correlation function 
have been reported in 
Refs.~\cite{Ohnishi:2016elb,Cho:2017dcy,Haidenbauer:2018jvl}, there is so far no work which takes account of 
all of these effects.

In the present Letter we investigate the $K^-p$ correlation function
by developing and using a proper coupled-channel framework.
Calculations are performed in the charge basis with six channels
($K^-p$, $\bar{K}^0n$, $\pi^-\Sigma^+$, $\pi^0\Sigma^0$, $\pi^+\Sigma^-$
and $\pi^0\Lambda$),
and the coupled-channel version of the correlation function
formula~\cite{Lednicky:1998r,Haidenbauer:2018jvl} is used.
Coulomb interactions between charged particles are treated consistently. 
The threshold energy differences among the various channels are taken into account when solving the coupled-channel Schr\"odinger equation.
In practice we have adopted the realistic $\bar{K}N$-$\pi\Sigma$-$\pi\Lambda$ coupled-channel
potential~\cite{Miyahara:2018onh}. This potential is constructed starting from chiral SU(3) dynamics~%
\cite{Hyodo:2011ur,IHW} and  constrained by fits to the existing $K^{-}p$ data. 
It will be demonstrated that 
the $K^-p$ correlation function recently measured 
by the ALICE Collaboration~\cite{Acharya:2019bsa}
is well explained by our calculations with reasonably tuned  source size ($R$)
and the source weight in the $\pi\Sigma$ channels ($\omega_{\pi\Sigma}$).
The effects of coupled channels are important in two ways:
the modification of the wave functions in the $K^-p$ channel,
and the conversion to $K^-p$
from $\bar{K}^0n$ and $\pi\Sigma$ generated from
the source functions in those channels.

\textit{Formalism:}
In high-energy heavy-ion collisions
and high-multiplicity events of $pp$ and $pA$ collisions,
the hadron production yields are well described by the statistical model. 
Under such conditions the correlations between outgoing particles are viewed as  generated
by the quantum mechanical scattering in the final state.

Consider two asymptotically observed particles, $a$ and $b$, with relative momentum $\bm{q} = (m_b\bm{p}_a - m_a\bm{p}_b)/(m_a+m_b)$. Let this two-particle state be fed by a set of coupled channels, each denoted by $j$.
In the pair rest frame of the two measured particles, 
their correlation function $C(\bm{q})$ is given as
~\cite{lednicky82:_influence,Haidenbauer:2018jvl}:
\begin{equation}
C(\bm{q})= \int d^3r \sum_j \omega_{j} S_j(\bm{r}) |\Psi^{(-)}_{j}(\bm{q};\bm{r})|^2
\ ,\label{Eq:KP}
\end{equation}
where the wave function $\Psi^{(-)}_{j}$ in the $j$-th channel is written as a function of the relative coordinate $\bm{r}$ in that channel, with outgoing boundary condition for the measured channel. Furthermore, 
$S_j(\bm{r})$ and $\omega_{j}$ are the (normalized) source function and its weight in the $j$-th channel.
The correlation function carries the information, through the wave functions $\Psi^{(-)}_{j}$, about the interactions in the channels $j$ contributing to the final state under consideration. 
One can extract this
information by properly determining the source function  (e.g., by combined fit to
data or constraints from other measurements) or by controlling the influence of the source
function (e.g., by varying the system size as advocated in Ref.~\cite{Morita:2016auo}).

We concentrate on the small $q=|\bm{q}|$ region and assume that only the $s$-wave part of the wave function is modified by the strong interaction.
The $j$-th channel component of the wave function with outgoing boundary condition in the $K^-p$ channel
(channel 1) is given as
\begin{equation}
\Psi^{(-)}_{j}(\bm{q};\bm{r})=[\phi^{C}(\bm{q};\bm{r})-\phi_{0}^C(qr)]\delta_{1j}
+ \psi_j^{(-)}(q;r)
,\label{Eq:wf1}
\end{equation}
where $r=|\bm{r}|$ and $\phi^{C}(\bm{q};\bm{r})$ is the free Coulomb wave function in the $K^{-}p$ channel,
$\phi^{C}_0(qr)$ is its $s$-wave component,
and $\psi_j^{(-)}(q;r)$ represents the $s$-wave 
function that includes both strong and Coulomb potential effects in the $j$-th channel. This wave function 
is subject to the outgoing boundary condition in the $K^-p$ channel as
specified below.

The wave function $\psi_j^{(-)}(q;r)$ is computed by solving the coupled-channel Schr\"odinger equation
\begin{eqnarray}
\mathcal{H} \psi(q;r)&=& E  \psi(q;r), \\
\psi(q;r)&=&^t\left[\psi_{1}(q;r),\psi_{2}(q;r), \cdots\right],
\end{eqnarray}
with Hamiltonian
\begin{eqnarray}
\mathcal{H}=\begin{pmatrix}
	-\frac{\nabla^2}{2\mu_{1}}+V_{11}(r) & V_{12}(r) & \dotsb \\
	V_{21}(r) & -\frac{\nabla^2}{2\mu_{2}}+V_{22}(r)+\Delta_{2} & \dotsb \\
	\vdots & \vdots & \ddots
\end{pmatrix},
\end{eqnarray}
where the channel indices $j=1,\dots, 6$ stand for $K^-p$, $\bar{K}^0n$, $\pi^-\Sigma^+$, $\pi^0\Sigma^0$, $\pi^+\Sigma^-$ and $\pi^0\Lambda$, in this order; $\mu_{j}$ and $\Delta_{j}$ represent the reduced mass in channel $j$ and the threshold energy difference relative to channel 1, respectively.
The diagonal potentials $V_{jj}(r)$ include the Coulomb term, $-\alpha/r$, in channels 1, 3, and 5 in addition to the strong interaction. The off-diagonal potentials are given by the strong interaction only. The momentum in channel $j$ is  $q_{j}=\sqrt{2\mu_{j} (E-\Delta_{j})}$ with $q\equiv q_{1}$ and $\Delta_{1}=0$. Through this relation, all the momenta $q_{j}$ can be expressed as functions of $q$.
The Schr\"odinger equation is then solved with the following boundary condition at $r\to\infty$:
\begin{equation}
\psi_j^{(-)}(q;r)\to \, {1\over {2iq_j}}
\left[\delta_{1j}\, {u_{j}^{(+)}(q_jr)\over r}
+A_{j}(q) \,{u_{j}^{(-)}(q_jr)\over r}\right]
\label{eq:BC}
\end{equation}
for open channels ($E>\Delta_{i}$), where $u_{j}^{(+)}$ and $u_{j}^{(-)}$ are the outgoing and incoming waves with a coefficient $A_{j}(q)$. In contrast to the standard scattering problem with normalized flux of the incoming wave in the incident channel, it is the outgoing wave in the measured channel that is normalized for the calculation of the correlation function. In the absence of the Coulomb interaction, $u_{j}^{(\pm)}(qr)/r=e^{{\pm}iqr}/r$ are spherical waves,
and the coefficients $A_{j}(q)$ are given by $\sqrt{(\mu_jq_j)/(\mu_1q_1)}
\,\mathcal{S}^\dagger_{1j}(q_{1}) $, with the $S$ matrix $\mathcal{S}_{ij}$. Including the Coulomb interaction we have
$u_{j}^{(\pm)}(qr)=\pm e^{\mp i\sigma_{j}}[iF(qr)\pm G(qr)]$
with $\sigma_{j}=\mathrm{arg}\Gamma(1+i\eta_{j})$, $\eta_{j}=-\mu_{j}\alpha/q_{j}$ with $F(qr)$ [$G(qr)$] being the regular (irregular) Coulomb function.
For closed channels ($E<\Delta_{i}$), the asymptotic form is given by substituting $q_j=-i\kappa_j=-i\sqrt{2\mu_j (\Delta_{j}-E)}$ as 
$\psi_{j}^{(-)}(r) \to A_{j}(q)\,u_{j}^{(-)}(-i\kappa_jr)/(2\kappa_{j}r)\propto e^{-\kappa_jr}/\kappa_jr$.
This is because the wave function component of the off-shell state can emerge only in the strong interaction region.
For spherically symmetric source functions the correlation function
can be written as
\begin{eqnarray}
C(q)=&
\int d^3r\,S_1(r)\left[|\phi^C(\bm{q};\bm{r})|^2-|\phi^C_0(qr)|^2\right]
\nonumber\\
&+4\pi\sum_j \int_0^\infty dr\,r^2\,\omega_{j}S_j(r)|\psi_j^{(-)}(q;r)|^2\,,
\label{Eq:CF}
\end{eqnarray}
where the left-hand side depends only on $q=|\bm{q}|$.
The normalization of the source function implies that the weight of the observed channel must be unity: $\omega_1 =1$~\cite{Lednicky:1998r}.

The $K^{-}p$ correlation function was calculated in Ref.~\cite{Cho:2017dcy} using the effective $\bar{K}N$ potential in Ref.~\cite{Miyahara:2015bya} within the model space of $K^{-}p$ and $\bar{K}^{0}n$ channels. 
Although the effects of the coupled $\pi\Sigma$ and $\pi\Lambda$ channels are implicitly included in the renormalized $\bar{K}N$ potential to reproduce the  scattering amplitude, the proper boundary condition~\eqref{eq:BC} was not imposed because the wave function does not contain explicit $\pi\Sigma$ and $\pi\Lambda$ components. The present calculation reduces to that in Ref.~\cite{Cho:2017dcy} when the channel couplings of $\bar{K}N\leftrightarrow\pi\Sigma,\pi\Lambda$ are switched off and the $\bar{K}^0n$ source function is ignored. It turns out, however, that there are sizable deviations of the present results from those in Ref.~\cite{Cho:2017dcy}. This indicates the importance of an explicit treatment of coupled channels in the $K^{-}p$ correlation function. 


We now
employ the wave functions 
in the full $\bar{K}N$-$\pi\Sigma$-$\pi\Lambda$ coupled-channel framework.
The starting point is chiral SU(3) dynamics at next-to-leading order ~\cite{IHW} which successfully describes the set of existing $K^{-}p$ scattering data together with the SIDDHARTA kaonic hydrogen data~%
\cite{Bazzi}.
An equivalent local
$\bar{K}N$-$\pi\Sigma$-$\pi\Lambda$ coupled-channel potential has been constructed to reproduce the corresponding scattering amplitudes~\cite{Miyahara:2018onh}.
Note that the coupled-channel effects contribute to the correlation function through the wave functions $\psi_j^{(-)}$ including $\psi_{K^-p}^{(-)}$.

\begin{figure}
\begin{center}
\includegraphics[width=0.5\textwidth]{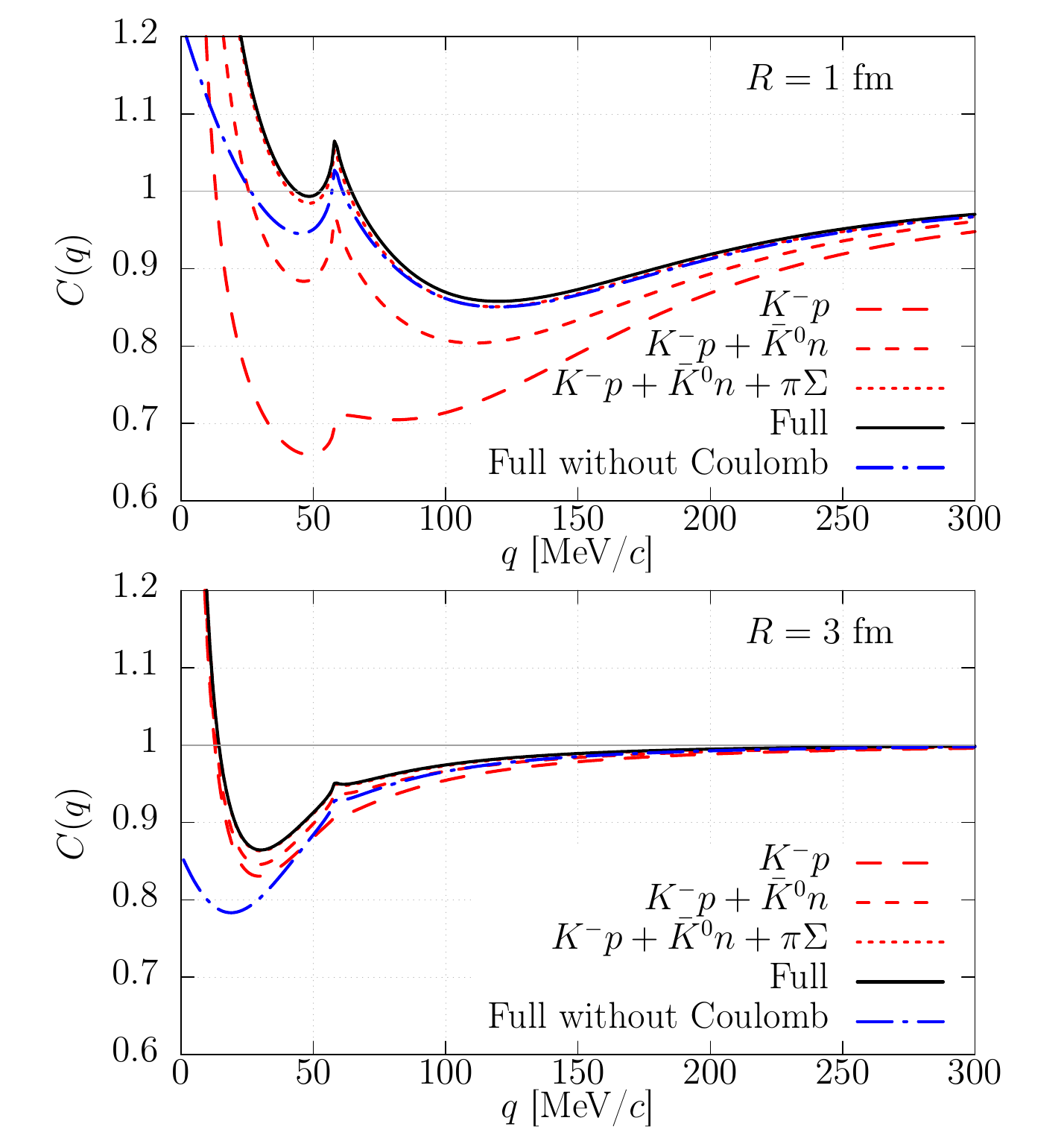}
\caption{$K^-p$ correlation function with $R=1$ fm (upper panel) and $R = 3~\mathrm{fm}$ (lower panel). 
The long-dashed line denotes the result with $K^-p$ component only. The short-dashed, dotted, and solid lines show the results in which the contributions from $\bar{K}^0n$, $\bar{K}^0n$ and $\pi\Sigma$, and from all coupled-channel components are added, respectively. The dash-dotted line denotes the full coupled-channel calculation without the Coulomb interaction. }\label{Fig:CFR}
\end{center}
\end{figure}

\textit{Results:}
The $K^-p$ correlation function and its breakdown into channels
are shown in Fig.~\ref{Fig:CFR} for source sizes of $R=1~\mathrm{fm}$ and $3~\mathrm{fm}$. 
We assume a common source function of Gaussian shape
for all channels,
$S_j(r)=S_R(r)\equiv \exp(-r^2/4R^2)/(4\pi R^2)^{3/2}$ with $\omega_j=1$.
For both source radii $R$ we can see the strong enhancement due to the Coulomb attraction 
at small momenta, demonstrated by comparison with the results omitting the Coulomb interaction. Also evident is
the cusp structure at the $\bar{K}^0n$ threshold at $q\simeq 58$ MeV/$c$.
Among the coupled-channel components,
the enhancement by the $\bar{K}^0n$ channel is found to be the largest,
and next in importance is $\pi\Sigma$.
The inclusion of the $\bar{K}^0n$ component also makes the cusp structure more prominent.
The $\pi^{0}\Lambda$ channel couples to $K^-p$ only in the $I=1$ sector; its effect is relatively weak.
Because the calculated wave functions in channels other than $K^-p$ have a sizable magnitude only at small distances, the contributions from  these components decrease with increasing source size.
This leads to a less pronounced cusp structure for the $R=3~\mathrm{fm}$ case.


Now we are prepared to compare the calculated $K^-p$ correlation function with data.
We allow for variations of the source size and weights,
which can be channel-dependent~\cite{Haidenbauer:2018jvl}.
Since a given source function with the weight
in the relative coordinate is 
obtained from a product of single-particle emission functions,
the weight should be proportional to the product of particle yields.
For example,
$\omega_{\pi^-\Sigma^+}/\omega_{K^-p}=N(\pi^-)N(\Sigma^+)/N(K^-)N(p)$.
The production yields $N(h)$ should be regarded as those of {\em promptly} 
emitted particles in order for those hadrons to couple into the final $K^-p$ channel. Those
primary yields are not directly observable. Thus,
we regard the source weights $\omega_j$ as parameters.
While the effect of the $\pi^0\Lambda$ channel is small and
the correlation function is not very sensitive
to $\omega_{\pi^{0}\Lambda}$, 
the effects of $\pi\Sigma$ channels are important because of the strong $\bar{K}N$-$\pi\Sigma$ coupling.
Then we fix $\omega_{\pi^{0}\Lambda}=1$ and vary the parameter $\omega_{\pi\Sigma}$ around the reference value,
obtained by the simplest statistical model estimate~\cite{Andronic:2005yp},
$\omega_{\pi\Sigma}^\mathrm{(stat)}\simeq \exp[(m_K+m_N-m_\pi-m_\Sigma)/T_c]\simeq 2.0$ with $T_c=154~\mathrm{MeV}$~\cite{Bazavov:2014pvz,Borsanyi:2013bia}.
As for the source size,
the ALICE collaboration fixed $R=1.18~\mathrm{fm}$
by assuming the same source size as that of $K^+p$, which was obtained by the femtoscopic correlation fit
based on the J\"ulich $K^+p$ interaction~\cite{Haidenbauer:2018jvl}, with Coulomb effects
treated by the Gamow factor correction.
Although this correction describes the Coulomb effect well 
for light systems such as $\pi$-$\pi$, it lacks the necessary accuracy for heavier systems~\cite{Morita:2016auo}.
Thus, we also consider the variation of $R$ in the fitting procedure.
While the source size can in principle be channel dependent, possible size differences between channels can be compensated by varying the source weights.
We therefore use a common source size in $\bar{K}N$, $\pi\Sigma$, and $\pi\Lambda$ channels.
We also assume that the source function has a Gaussian shape
and the source weight is isospin symmetric.

\begin{figure}[htbp]
	\begin{center}
		\includegraphics[width=0.5\textwidth]{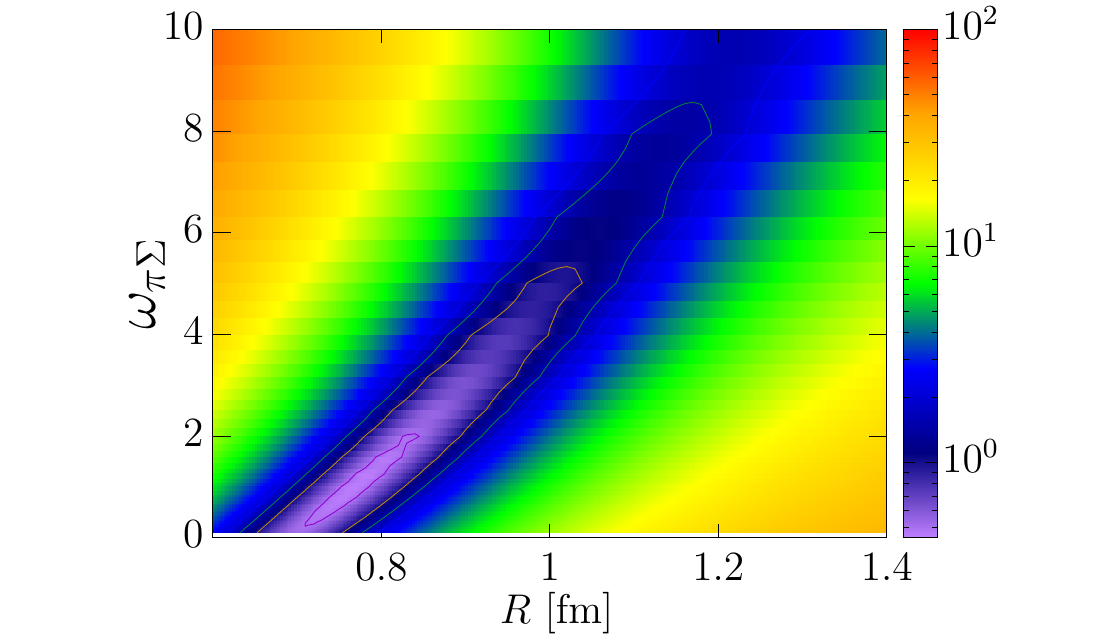}
	\end{center}
	\caption{Reduced $\chi^2$ distribution in the $(R,\omega_{\pi\Sigma})$ plane. From inward out the contour lines correspond to $\chi^2/ \mathrm{d.o.f.}= 0.5, 1, 1.5$ and 2, respectively.
	}\label{Fig:chi_wR}
\end{figure}

The measured correlation function is assumed to be described in the form~\cite{Acharya:2019bsa}
\begin{equation}
C_\mathrm{fit}(q)=\mathcal{N}\left[1+\lambda\left\{C(q)-1\right\}\right],\label{Eq:fit}
\end{equation}
where $\mathcal{N}$ is a normalization constant and
$\lambda$ is the pair purity parameter, known also as the chaoticity parameter.
The pair purity parameter is experimentally determined through a Monte Carlo simulation, $\lambda_\mathrm{exp}=0.64\pm 0.06$, so we allow for variations of $\lambda$ within $1\sigma$.
We fit the correlation function
data in the momentum range $q<120~\mathrm{MeV}/c$,
where the distortion of the $s$ wave is considered
to give the dominant contribution.

In Fig.~\ref{Fig:chi_wR} the $\chi^2 / \mathrm{d.o.f.}$ distribution is plotted in the $(R,\omega_{\pi\Sigma})$ plane.
A good fit ($\chi^2/\mathrm{d.o.f.} \lesssim 1$) is achieved 
in the region from $(R,\omega_{\pi\Sigma})=(0.6~\mathrm{fm},0)$
to $(1.1~\mathrm{fm},5.0)$.
The source size $R \simeq 1~\mathrm{fm}$ is reasonable for $pp$ collisions,
while $\omega_{\pi\Sigma}$ should be consistent
with the simple statistical model estimate within a factor of 2 to 3.
Thus, we consider parameter sets in this region 
with $0.5\leq \omega_{\pi\Sigma} \leq 5$ as equally acceptable.
On the other hand, if we take the $R=1.18~\mathrm{fm}$ as adopted by the ALICE Collaboration,
$\omega_{\pi\Sigma}\gtrsim 8$ gives a good fit, 
but such large $\omega_{\pi\Sigma}$ values appear to be significantly beyond the statistical model estimate.

\begin{figure}[htbp]
	\begin{center}
		\includegraphics[width=0.5\textwidth]{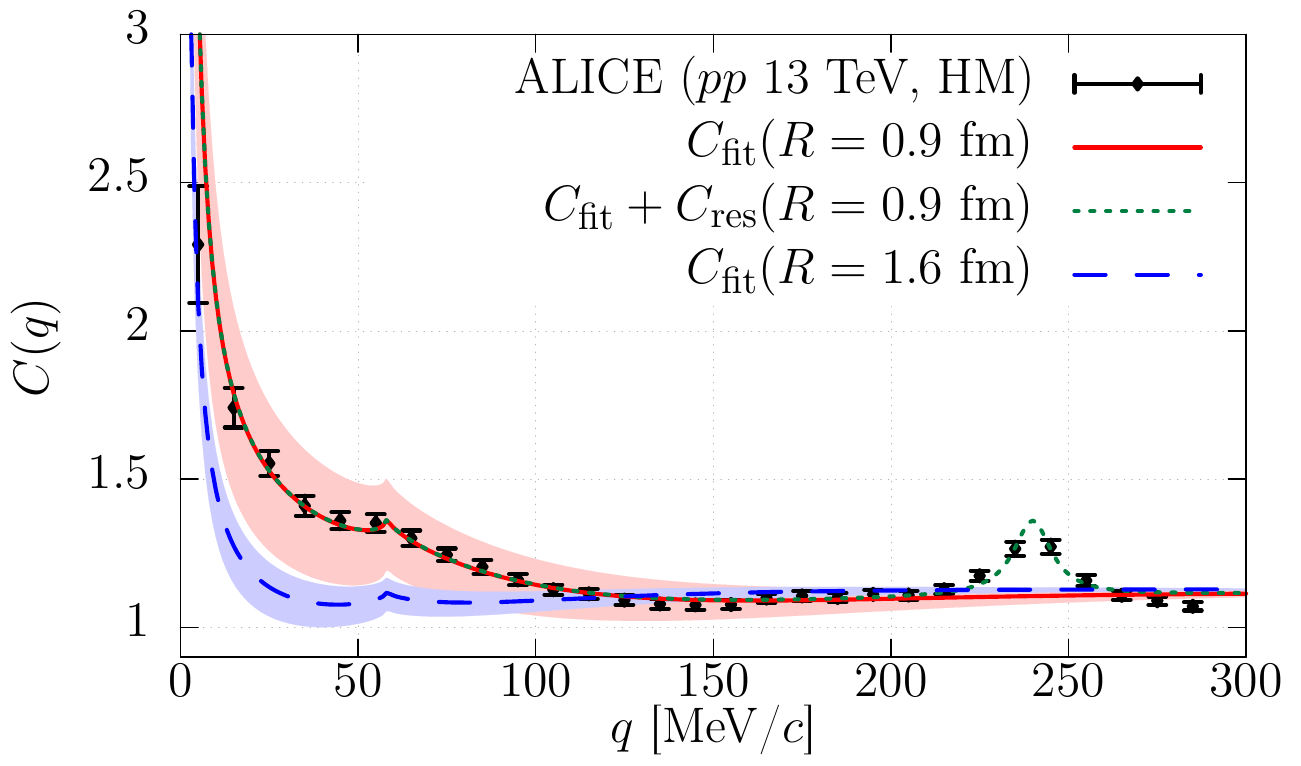}
	\end{center}
	\caption{Correlation function with the best fit parameters 
(solid line). The result including the $\Lambda(1520)$ contribution is shown by the dotted line. The dashed line shows the prediction with $R=1.6 \ \mathrm{fm}$. Its shaded area shows the uncertainty with respect to the variation of $\omega_{\pi\Sigma}$. For comparison, we also plot the corresponding area for the case with $R=0.9 \ \mathrm{fm}$.  The ALICE data set is taken from Ref.~\cite{Acharya:2019bsa}.
	}\label{Fig:CFbestfit}
\end{figure}

Figure~\ref{Fig:CFbestfit} shows the fitted $K^-p$ correlation function with $R = 0.9~\mathrm{fm}$ as an example of a result satisfying $\chi^2/\mathrm{d.o.f.} < 1 $. The other parameters are chosen as
\begin{equation}
\quad \omega_{\pi\Sigma} = 2.95,  \quad \mathcal{N} = 1.13,\quad  \lambda = 0.58, \label{Eq:params}
\end{equation}
to give the minimum value 
of $\chi^2/\mathrm{d.o.f.} =0.58$.
The enhancement in the low-momentum range and the characteristic cusp structure are evidently well reproduced. 
Recalling the importance of the $\pi\Sigma$ component in the $K^-p$ correlation as shown in Fig.~\ref{Fig:CFR}, the sizable value of $\omega_{\pi\Sigma}$ indicates that the contribution from the $\pi\Sigma$ source is essential to reproduce the data.


The peak structure seen in Fig.~\ref{Fig:CFbestfit} around $q\sim 240$ MeV/$c$ represents the $\Lambda(1520)$ resonance.
The contribution from this resonance can be simulated by a Breit-Wigner function: 
\begin{equation}
C_\mathrm{res}(q)=\frac{b\Gamma^2}{(q^{2}/2\mu_{K^-p}+m_p+m_{K^-} -E_R)^2+\Gamma^2/4}, \label{Eq:Lambda1520}
\end{equation}
with parameters $b$, $E_R$, and $\Gamma$.
We can isolate the resonance by subtracting $C_\mathrm{fit}(q)$ from the correlation data, using the parameters of Eq.~\eqref{Eq:params} and $R=0.9~\mathrm{fm}$. The remaining structure in the interval  $150~\mathrm{MeV}/c<q<300~\mathrm{MeV}/c$ is then fitted by Eq.~\eqref{Eq:Lambda1520}.
The resulting values of the resonance parameters are $E_R = 1520.9~\mathrm{MeV}$ and $\Gamma=9.7~\mathrm{MeV}$, consistent with the mass $M_{\Lambda(1520)}=1517 \pm 4~ \mathrm{MeV}$
and width $\Gamma_{\Lambda(1520)}= 15^{+10}_{-8}~\mathrm{MeV}$ of $\Lambda(1520)$ listed in Ref.~\cite{Tanabashi:2018oca}.
As shown in Fig.~\ref{Fig:CFbestfit}, the sum of $C_{\mathrm{fit}}(q) $ and  $C_\mathrm{res}(q)$ reproduces the peak at $q\sim 240~\mathrm{MeV}$ very well.

Finally we give predictions for the $K^-p$ correlation function if extracted from larger systems.
In $pA$ and $AA$ collisions the source size is expected to be larger than the one in $pp$ collisions:
$R=(1$-$2)~\mathrm{fm}$ for high-multiplicity events in $pA$ collisions
and $R=(2$-$5)~\mathrm{fm}$ in $AA$ collisions.
In Fig.~\ref{Fig:CFbestfit}, we show the theoretical correlation function, Eq.~\eqref{Eq:fit}, at a system size of $R=1.6~\mathrm{fm}$, using the same parameter set as before, Eq.~\eqref{Eq:params}, for demonstration.
In order to estimate the uncertainty coming from the less well known $\omega_{\pi\Sigma}$, we vary its value between $0.5$ and $5.0$. 
One expects that when the source size is increased, the enhancement of the correlation function is limited to the small $q$ region and the $\bar{K}^0n$ cusp becomes less pronounced.
The sensitivity to $\omega_{\pi\Sigma}$ 
is weaker for the larger systems,
for which the contribution from the $\pi\Sigma$ source is less important.


\textit{Summary:}
The $K^-p$ femtoscopic correlation function has been analysed
using the realistic coupled-channel potential of Ref.~\cite{Miyahara:2018onh}.
This potential is constructed to reproduce the amplitudes resulting from next-to-leading order chiral SU(3) dynamics~%
\cite{IHW}.
Based on the coupled-channel correlation function
formula~\cite{Lednicky:1998r,Haidenbauer:2018jvl},
we have developed a scheme to calculate the correlation function consistently including all effects of coupled channels, Coulomb potential and threshold differences in the individual channels. 
The coupled channels play an important role, enhancing the correlation function and producing a prominent threshold cusp effect.
The $K^-p$ correlation function data obtained by the ALICE collaboration~\cite{Acharya:2019bsa} are well reproduced. 
The allowed range of values for the source function weight, $\omega_{\pi\Sigma}$, of the $\pi\Sigma$ channel is roughly consistent with a statistical model estimate. 


We have also presented a prediction of the $K^-p$ correlation function
for a generic larger system. 
In this case the driving coupled channels become less important as the source size increases. 
Analyzing correlation data 
extracted from collisions of larger systems
could provide additional systematics for probing and constraining the $K^-p$ amplitude in low-energy regions not accessible by scattering experiments. 

While the $\bar{K}N$-$\pi\Sigma$-$\pi\Lambda$ coupled-channel approach \cite{Miyahara:2018onh,Hyodo:2011ur,IHW}
has successfully passed the present femtoscopy test (though with adjustment of source parameters), it will be useful also to examine alternative models of low-energy $K^-p$ interactions in order to gain more systematic insights into the capabilities of such correlation function studies.
\begin{acknowledgments}
The authors thank
Laura Fabbietti, 
Valentina Mantovani Sarti, 
Ot\'on V\'azquez Doce, 
Johann Haidenbauer, 
and other participants of the YITP workshop (YITP-T-18-07)
for useful discussions.
Y.K., T.H., and A.O. gratefully acknowledge the hospitality
of the members in Technical University of Munich
during their stay in Munich.
This work is supported in part by the Grants-in-Aid for Scientific Research
from JSPS (Nos.
19H05151, 
19H05150, 
19H01898, 
and 
16K17694), 
by the Yukawa International Program for Quark-hadron Sciences (YIPQS),
by the Polish National Science Center NCN under Maestro Grant No.
EC-2013/10/A/ST2/00106, 
by the National Natural Science Foundation of China (NSFC)
and the Deutsche Forschungsgemeinschaft (DFG) through the funds
provided to the Sino-German Collaborative Research Center
``Symmetries and the Emergence of Structure in QCD"
(NSFC Grant No. 11621131001, DFG Grant No. TRR110),
by the NSFC under Grants No. 11835015 and No. 11947302,
by the Chinese Academy of Sciences (CAS)
under Grant No. QYZDB-SSW-SYS013 and No. XDPB09,
and  the CAS President's International Fellowship Initiative (PIFI) under Grant No. 2020PM0020.
\end{acknowledgments}

%
\end{document}